\documentclass[preprint2]{aastex}

\usepackage{amsmath} 
\usepackage{amssymb} 
\usepackage{graphicx}
\usepackage{textcomp}
\usepackage{amsthm}
\usepackage{gensymb}
\usepackage{natbib}

\def\that{{\hat t}}

\def\msun{M_\odot}
\def\rsun{R_\odot}
\def\ten#1{\times 10^{#1}}
\def\aten#1{10^{#1}}
\def\VEV#1{\left\langle #1\right\rangle}

\def\mn{{m_{9}}}

\slugcomment{ }

\shorttitle{Experimental Limits on Primordial Black Hole Dark Matter}

\begin{document}


\title{Experimental Limits on Primordial Black Hole Dark Matter from the First Two Years of Kepler Data}


\author{Kim Griest, Agnieszka M. Cieplak} 
\affil{Department of Physics, University of California,
    San Diego, CA 92093}
\author{Matthew J. Lehner}
\affil{Institute of Astronomy and Astrophysics,
    Academia Sinica. P.O. Box 23-141, Taipei 106, Taiwan}
\affil{Department of Physics and Astronomy,
        University of Pennsylvania, Philadelphia, PA 19104\\ Draft: 3 June 2013}



\begin{abstract}

We present the analysis on our new limits of the dark matter (DM) halo consisting of primordial black holes (PBHs) or massive compact halo objects (MACHOs). We present a search of the first two years of publicly available Kepler mission data for potential signatures of gravitational microlensing caused by these objects, as well as an extensive analysis of the astrophysical sources of background error. These include variable stars, flare events, and comets or asteroids which are moving through the Kepler field. We discuss the potential of detecting comets using the Kepler lightcurves, presenting measurements of two known comets and one unidentified object, most likely an asteroid or comet. After removing the background events with statistical cuts, we find no microlensing candidates. We therefore present our Monte Carlo efficiency calculation in order to constrain the PBH DM with masses in the range of  $2\ten{-9}\msun$ to $\aten{-7}\msun$ . We find that PBHs in this mass range cannot make up the entirety of the DM, thus closing a full order of magnitude in the allowed mass range for PBH DM.

\end{abstract}





\section{Introduction}
The nature of the dark matter (DM) remains one of the most important unsolved 
problems in science (Feng 2010 and references therein).  We know that it has a universal 
density around 5 times larger than that of material made of 
ordinary atoms, and is an essential 
ingredient of the current consensus cosmological model (Ade, et al. 2013), 
but have little information as to its actual nature.  Hundreds of candidates 
have been proposed, with the most popular being
various candidates from beyond the standard model of particle physics, many 
involving the lightest supersymmetric particle (LSP).  Besides the 
thousands of theoretical papers on Supersymmetric (SUSY) weakly
interacting massive particle (WIMP) candidates, there have been many 
experimental searches for these DM candidates (Feng 2010).  
Despite some tantalizing hints, currently there seems to 
be no compelling WIMP candidate, SUSY or otherwise.  The lack of detection of 
SUSY partners of any sort from the recent Large Hadron Collider (LHC) run is 
especially disappointing in this 
regard (Chatrchyan et al. 2012; ATLAS Collaboration 2013; 
CMS Collaboration 2013).
SUSY has been so popular over the past few decades because
it seemed to simultaneously solve the hierarchy fine-tuning 
problem (Martin 2011) and give a ``natural" DM candidate.  
If SUSY partners existed below the TeV scale then the LSP annihilation 
cross section tended to be in the range which gave the measured relic 
abundance of LSP particles (Jungman et al. 1996).  This so-called WIMP 
miracle motivated much theoretical and experimental work on SUSY DM.
The discovery of the Higgs Boson with a mass around 126 
GeV (Aad, et al. 2013; Chatrchyan et al. 2013),
along with the lack of SUSY particles below the TeV 
scale removes some of this motivation.  
The LSP can still be the DM, but now some fine-tuning will be required to arrive at the measured relic abundance.  

More generally, we note that the theoretical and experimental emphasis on the admittedly 
elegant SUSY models over the past few decades may have been misplaced.  
Many important experimental discoveries have not verified our aesthetic desire for simple unified models.  In fact, 
experimental breakthroughs have led us in almost the opposite direction.  The discovery of cosmic microwave background (CMB) anisotropies  points
strongly towards an epoch of cosmic inflation in the early universe, most likely caused by a (finely tuned) inflaton scalar field.  The discovery of
the dark energy points to an extremely finely-tuned cosmological constant, or quintessence-like scalar field.  (Even a cosmological constant can
be thought of as the vacuum expectation value of a scalar field).  
Finally, the Higgs Boson mass of around 126 
GeV (Aad, et al. 2013; Chatrchyan et al. 2013)
seems to require some fine tuning.  

Thus, perhaps we should abandon our Occam's razor proclivities and accept that finely-tuned scalar fields seem to be
part of modern physics and thus may also be part of the solution to the DM problem.  
If so, then primordial black holes (PBH) should be seriously considered as DM candidates.

In light of the above, PBH DM has several things going for it.  First it is one of the few standard model
DM candidates.  No need for SUSY or superstring inspired Grand Unified Models.
The DM problem is detected primarily through gravity, so PBH DM would be a gravitational
solution to a gravitational problem.  There are many ways to create PBH DM, and many of these
involved finely-tuned scalar fields.  In the past,
this has been taken as a negative for PBH DM,
but the above considerations may allow rethinking of this attitude.
For example, there are several
double inflation models where one inflation solves the flatness, etc. problems that inflation
is invoked to solve and the other inflationary epoch gives rise to 
PBHs (Frampton, et al. 2010; Kawasaki, Sugiyama, \& Yanagida 1998)
which can then become
the dark matter.  Recall that one of the initial motivations for the scale-free Harrison-Zeldovich
spectrum of primordial fluctuations was to avoid creating PBHs.
Thus PBHs can easily be made via a tilted spectrum of fluctuations or through
production of particles that then create the black holes (BH).   

Note that some mechanisms
of PBH creation result in a broad spectrum of PBH masses, but that double inflation mechanisms
tend to create a nearly delta-function spectrum with the masses strongly concentrated
near the mass enclosed in the horizon at the epoch of formation.  Also note
that there have been dozens of other suggested ways 
to create PBH DM (see, for example, Khlopov 2008; Frampton, et al. 2010; Carr, et al. 2010). 
Finally note that if PBHs are created early enough and in appropriate mass ranges
they can evade big bang nucleosynthesis (BBN) and CMB constraints and make up the entirety of the DM.

\section{Gravitational Microlensing of Kepler Satellite Data}

Since the work of Paczynski (1986) gravitational microlensing has
been used as a powerful method to probe the DM in the Milky Way. If the DM is composed of massive compact halo objects (MACHOs), its signature could be detected through the occasional magnification of stellar flux, when the objects pass near the line-of-sight to a star, as they move through the Milky Way halo.
Many theoretical and experimental results
have been obtained which together have eliminated MACHO DM (which includes PBH DM)
in the mass range from $3\ten{-8}\msun$ to $30\msun$ from being the entirety of the 
DM (Carr, et al. 2010; Tisserand, et al. 2007; Alcock, et al. 2001, Alcock, et al. 1998; 
Alcock, et a. 1996; Griest 1991).
The strongest limits above on MACHO DM come from observing
programs toward the Large Magellanic Cloud,  but here we follow
Griest, et al. 2011 (Paper I) and Cieplak \& Griest, 2013 (Paper II) in using microlensing of the nearby
Kepler mission source stars to search for potential MACHO DM signatures. First constraints of PBH DM are being presented in Griest, et al. 2013. Here, we present a full-scale analysis of these constraints, including a more extensive study of the background sources of error.
Throughout this paper we follow the methods developed by the authors mentioned
above.

The Kepler telescope has a 1 m aperture with a $115$ deg$^2$ 
field-of-view and is in an Earth trailing heliocentric 
orbit (see Koch et al. (2010) and Borucki, et al. (2010) for a description of the Kepler 
mission). It takes photometric measurements of around 150,000 stars every 
30 minutes towards the Cygnus-Lyra region of the sky. 
The telescope was launched in March 2009. 
The main goal of the Kepler mission is to discover extra-solar planets by
the transit technique.  For well aligned systems, a planet will cross in
front of the stellar limb and cause the measured stellar flux to drop by
a small amount.  In order to detect Earth size planets, Kepler has exquisite
photometric accuracy, measuring fluxes to one part in 10,000 or better. 
In this paper we analyze these same stellar lightcurves, but look instead
for short duration increases in stellar flux caused by gravitational 
microlensing of a PBH as it passes near the line-of-sight of the star.

Naively, the nearby Kepler source stars should not be very useful for microlensing,
since they are at a typical distance of 1 kpc and only 150,000 in number.
In general, the sensitivity of microlensing searches for dark matter is
proportional to the distance to the stars, the number of stars monitored, and
the duration of the observing program (Paczynski 1986). 
Previous microlensing searches for 
DM towards the Large Magellanic Cloud (Alcock, et al. 2000; Tisserand et al. 2007)
monitored more than 12 million stars at a distance of
50 kpc for 8 years, naively giving a factor of several thousand larger
sensitivity than the Kepler source stars.  However, as shown in 
Papers I and II, the finite size of the Kepler stars
coupled with Kepler's high precision photometry imply that Kepler is more
sensitive than any previous microlensing experiment for PBHs in the
mass range $2\ten{-10}\msun$ to $2\ten{-6}\msun$.

\section{Data Analysis and Event Selection}
Here we present a preliminary analysis of 2 years of publicly 
available Kepler data (quarters 2 through 9) (http://archive.stsci.edu/kepler, 2013;
Fraquelli \& Thompson 2011); 

Each quarter, the Kepler team releases around 150,000 lightcurve
files each containing around 4400 flux measurements taken over 
about 90 days, with a cadence of around 30 minutes (more precisely: 29.425 minutes).
The data we use consists of the reported time of each flux measurement ({\bf time}), 
the photmetric flux ({\bf pdcflux}), the flux error ({\bf pdcfluxerr}), and the quality flag 
({\bf sapquality}).
We use the Kepler pipeline flux data {\bf pdcflux} which removes various trends from 
the data.  The aperture photometry fluxes are also available but we do not use
these.
Since the Kepler team is looking for dips in the flux which last only a few
hours and we are looking for microlensing bumps in the data which last a
similarly short time, the detrending procedure the Kepler team applies to
the data should be good for us as well.
The quality flag is set non-zero if problems exist in the photometry, for
example cosmic rays, reaction wheel desaturation, etc. (Fraquelly \& Thompson 2011).
We ignore all data where the quality flag is non-zero.

We divide the good fluxes by the mean flux of the lightcurve, which is
calculated over 300 data points towards the center of the lightcurve, and subtract unity from 
this quantity to produce a fractional magnification lightcurve.
We divide the flux errors by the mean as well to give us appropriate errors on
the fractional magnifications.
These lightcurves
are used for our initial search, however, when performing fits to microlensing shapes, etc.
we renormalize our selected lightcurves using the median flux of the whole lightcurve, which gives
a more robust baseline.
For each lightcurve we calculate a set of statistics that allow us to
identify microlensing candidates events and also to distinguish various types of background
events.  These statistics and the selection criteria are listed in 
Table~\ref{tab:cuts}.

\begin{deluxetable}{ll}
\tabletypesize{\scriptsize}
\tablecaption{Definitions of Statistics and Selection Criteria\label{tab:cuts}}
\tablewidth{0pt}
\tablehead{
\colhead{\bf Statistic} &
\colhead{\bf Definition}} 
\startdata
$\VEV{f}$ & average of $f_i$ over all good measurements in lightcurve \\ 
$A_i$ & flux$_i/\VEV{\rm flux}$ \\
$\sigma_i$ & reported error of flux normalized by average flux\\
bump & sequence of 4 or more contiguous fluxes with $A_i-1 \geq 3\sigma_i$\\
nbump & number of bumps in lightcurve\\ 
bumplen & number of contiguous fluxes with $A_i-1\geq 3 \sigma_i$\\ 
lag1autocorr & ${1\over N}\sum_{i=1}^{i=N}((A_i-1) (A_{i+1}-1))$\\
bumpvar & $\sum |A_i-1|/\sigma_i$ over points under bump\\
leftedgevar &$\sum |A_i-1|/\sigma_i$  over 2 bumplen points starting 6 bumplen before bump\\
rightedgevar &$\sum |A_i-1|/\sigma_i$  over 2 bumplen points starting 4 bumplen after bump\\
dof & number of data points within 5 bumplen of peak minus number of fit parameters \\
mlchi2dof & $\chi^2$ of fit to microlensing shape divided by dof \\
fchi2dof & $\chi^2$ of fit to exponential flare shape divided by dof \\
chi2in & $\chi^2$ of microlensing fit for points with time, $t_i$, such that
$t_0 - 1.5 \that < t_i < t_0+1.5\that$\\
chi2out & $\chi^2$ of microlensing fit for points with time, 
$t_0 - 6 \that < t_i < t_0+6\that$, but not in chi2in\\
$Nasy$ & number of points near peak time, $t_0$, for asymmetry; larger of
$1.5 \lambda$ and $2\that$ \\
asymmetry &  $\sum_{ Nasy\ \rm points} |A(t_0-t_i) - A(t_0+t_i) |/
(\sum A(t_i) - Nasy A_{min})$ \\

\tableline
\tableline
{\bf Selection Criterion} & {\bf Purpose}\\
\tableline
0 $<$ nbump $<$ 3 & remove variable stars and stars with no transient \\
bumplen $\geq$ 4 & level 1 trigger (significant bump) \\
bumplen $\geq$ 5 & remove short duration flare events\\
lag1autocorr $>$ 0.7 & remove obvious variable stars \\
bumpvar $> {1\over2}5.5($ leftedgevar + rightedgevar$)$ & signal to noise cut when reported errors are non-Gaussian\\
edgecriterion $>$ 0 & remove bumps that start or end in bad data \\
mlchi2dof $<$ 0.75 fchi2dof & microlensing fit significantly better than flare fit \\
mlchi2dof $<$ 3.5 & microlensing fit is not too bad\\
asymmetry $<$ 0.17 & remove short duration flare events\\
chi2in/chi2out $<$ 4 & remove events where
$\chi^2_{\rm dof}$ under peak is much 
worse than $\chi^2_{\rm dof}$ outside peak area\\
\enddata
\end{deluxetable}

\subsection{Statistics and Selection Criteria}
Since we are looking for very low magnification, short duration bumps in a large amount
of data the main backgrounds are measurement noise and stellar variability.
Our level 1 requirement is that a candidate event contain a ``bump",
defined as 4 sequential
flux measurements that are 3-standard deviations higher than the mean
flux.  (This was also the criterion applied in our theoretical papers: Paper I and Paper II.)

Naively applied, this criteria selects about 50\% of all the 150,000
lightcurves.  The problem is that given the extreme precision of Kepler
photometry, around 25\% of main sequence stars and 95\% of giant stars have
measurable variability (Ciardi, et al. 2011).
So before applying our bump selection criteria we first identify and remove
``variable stars" from the data base.  We call a star a ``variable star"
if it contains more than 2 bumps (defined above) and if the autocorrelation
function of the lightcurve calculated with a lag time of one measurement (around 30 minutes)
is larger than 0.7.  This {\bf lag1autocorr} statistic is a measure of
how correlated one flux measurement is with the next.  Since most variable
stars vary on time scales much longer than 30 minutes, we expect variable
stars to have a large value of this statistic.
The threshold value of 0.7 is set by requiring that the simulated microlensing
events we add to lightcurves to measure our detection efficiency are not 
removed by this cut.  In general, the cut values of each selection criterion
below is set as a compromise.  We want to  select as many of our added
simulated microlensing events as possible, while still eliminating whatever
background the cut is intended to remove.  Thus our cuts are as loose as
possible while eliminating background.
All of our cuts are listed and defined in Table~\ref{tab:cuts}.

Using these two criteria we find that we remove about 34\% of the dwarf
stars and around 91\% of the giant stars from further consideration.
Note that these numbers are consistent with the results of Ciardi, et al. (2011)
mentioned above, whose numbers we used in Paper II for our theoretical estimate.
Note that we are using the term ``giant star" very roughly to mean
any star with radius greater than 3 $\rsun$, as listed in the Kepler lightcurve file.

After eliminating variable stars our bump selection criteria gives us
around 10,000 candidate events.  However,
the Gaussian error approximation implied by our bump criterion is not
adequate, so we calculate three local measures of variability for
each bump we find:  {\bf bumpvar}, which is the absolute value 
of fractional flux increase divided by the error 
averaged over the duration of the bump
({\bf bumpvar} $= \VEV{|A-1|/\sigma}$); 
{\bf leftedgevar} which is the same for
the region of the lightcurve that occurred just before the time of the bump,
and {\bf rightedgevar} which is the same for a small portion of the lightcurve
just after the bump occurred.  We then demand that the variation during
the time of the bump is 5.5 times more significant than the variation 
that occurred just before and after the bump.  
This selection criterion eliminates bumps in lightcurves caused by
a sequence of very noisy measurements, where the noise is non-Gaussian, meaning 
the reported error-bars are not reliably determining the probability of outlier fluxes.

After applying these criteria we still find a large number of candidate
bumps.  Many of these bumps happen at the same time and have the same shape
across a large number of lightcurves.  These are most likely caused by
systematic problems in the photometry coming from either instrumental 
problems in the Kepler satellite or 
the detrending software.  We especially notice these systematic error
bumps just before or after regions of ``bad data", areas of lightcuves which have
the quality flag set.  To deal with these problems we compiled ranges
of dates over the 8 quarters which contained substantial numbers of
similarly shaped bumps.  These ranges of times are organized in a ``bad data"
file and are removed from all the lightcurves before calculating any statistics.

\begin{figure}[htb!]
\begin{center}
\includegraphics[scale=.4,trim=0.5in 0 0 0]{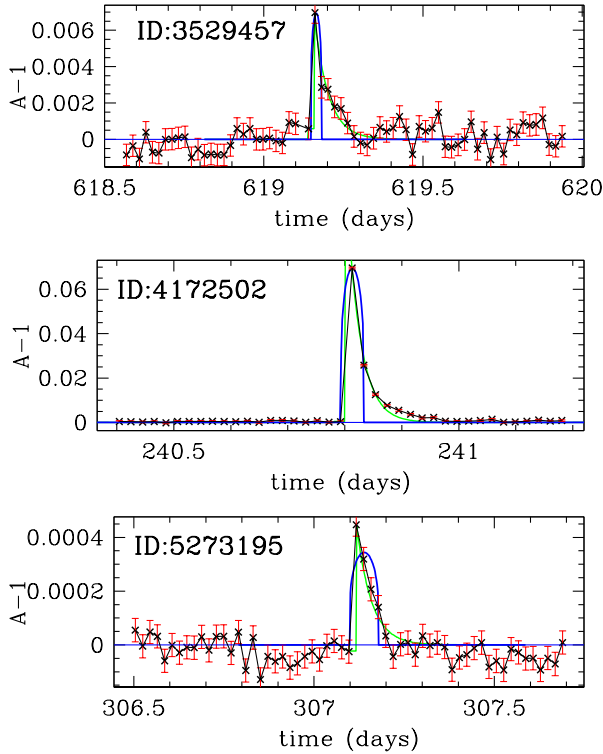}
\caption{
Examples of bumps in Kepler lightcurves caused by stellar flares.
The Kepler source star IDs are shown in the upper left corners, 
the solid blue line is the rather poor fit to the microlensing shape, and the solid green line is a (better) fit to flare event.
The top panel shows a medium amplitude flare event, the middle panel shows a high amplitude
event, and the bottom panel shows a very low amplitude flare.} 
\label{fig:flarelc}
\end{center}
\end{figure}

\begin{figure}[htb!]
\begin{center}
\includegraphics[scale=.4,trim=0.5in 0 0 0]{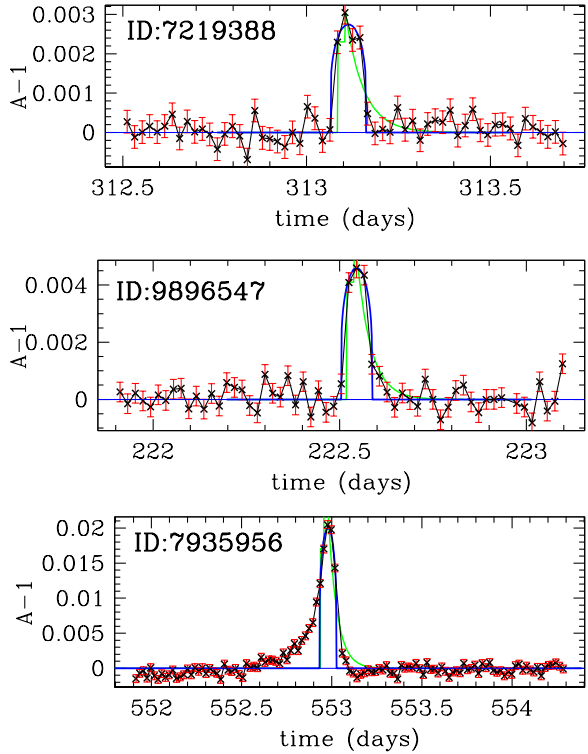}
\caption{
Examples of bumps in Kepler lightcurves.  The top two panels show very short duration events
that can be well fit with either microlelensing or flare shapes.  The bottom panel
is an event of unknown origin that is a poor fit to the microlensing shape.
The Kepler source star IDs are shown as are fits 
to microlensing (solid blue line) and flare (solid green line) shapes.
\label{fig:badlc} }
\end{center}
\end{figure}

After removing the bad data, the variables, and the noise bumps as discussed
above we still have around 100 candidate bumps per quarter,
some still caused by noise but many caused by short term variability.
Some example lightcurves are shown in 
Figures~\ref{fig:flarelc},\ref{fig:badlc},\ref{fig:cometlc5}, and \ref{fig:cometlc9}.
Examination of the lightcurves show that the selected bumps 
come in several categories.  
First, there are dozens of bumps that start during periods of ``bad data"
and then exponentially approach the mean flux over several hours or days.
To eliminate these we apply an {\bf edgecriterion}, demanding that our best
fit microlensing shape not start or end in a region of ``bad data" or
off the edge of the data; that is, we demand that the entire bump be contained
in the good data.

Next, the largest category of bumps by far, containing a
few hundred events, are lightcurves with
a highly asymmetric bump that we determined is probably due 
to a stellar flare
caused by magnetic activity in the stellar atmosphere.  
Examples are shown in Figure~\ref{fig:flarelc}.
To eliminate these flare events 
we compare two non-linear fits to each bump, the first using
the expected microlensing shape 
given by formulas in Cieplak \& Griest (2013)
\footnote{In fact, to speed the processing we fit only a linear limb darkened shape
(Equation 11 of Paper II) and not the full microlensing lightcurve shape.  
For the PBH masses considered, the small size of the Einstein Ring relative
to the projected size of the stellar limb means this is a good approximation.
If we find good microlensing candidates we would, of course, fit with the
complete microlensing profile.}
and the second using a flare event shape, 
which has a nearly instantaneous rise followed
by an exponential drop back to the median lightcurve flux.
For microlensing there are four parameters that determine the lightcurve shape:
$t_0$, the time of peak (time of closest approach between the lens and source), $\hat t$, the duration of the
event, $A_{max}$, the peak magnification, and $\epsilon_{min}$, 
the distance of closest approach between lens and source lines-of-sights divided by the stellar radius in the lens plane.
For the flare shape, we have $t_0$, the peak time,
$A_{max}$, the peak amplitude, and $\lambda$,
the exponential decay constant.  While we are assuming the rise is 
instantaneous and is followed by an exponential decay,
the rise can occur at any time during the 30 minute Kepler flux integration. 
So the flux measurement immediately preceding the measured peak flux
can have any value between the baseline and the peak value.  
Thus we add another degree-of-freedom (dof) to
the flare fit which is the value of the flux at the time
immediately preceding the peak time.
Examples of these fits are shown in 
Figures~\ref{fig:flarelc} and \ref{fig:badlc}, where we show the data as well 
as both the microlensing and flare fits. 


In order to eliminate the flare events
we then require that: {\bf mlchi2dof} $< 0.75 \times ${\bf fchi2dof}, where
{\bf mlchi2dof} is the chi-squared per degree-of-freedom (dof) of the microlensing fit,
and {\bf fchi2dof} is chi-squared per dof of the flare star fit.  We fit
over a region of the lightcurve up to ten times longer than the bump length,
where the bump length is defined as the number of 3-sigma high flux points
we used for the level 1 trigger.
This selection criteria works extremely well in removing 
longer duration (i.e. many hour) flare events, but for candidate events
lasting only a few hours does not always remove the flare event.
See Figure~\ref{fig:flarelc} for longer duration events where the fits
clearly distinguish microlensing from flaring, and 
the top two lightcurves in Figure~\ref{fig:badlc} where both 
microlensing and flare fits are
satisfactory due to the small number of data points in the bump.
The problem is that there is a large background of flare events which
have a wide range of durations.  We therefore expect many short duration
flare events with only a few points in the bump.  However, our fits lose
their power to distinguish flares from microlensing when the 
number of data points is small.
In addition, we start the fit for the flare shape 
from the highest point, but allow one previous flux to be high due to
the roughly 30 minute Kepler integration time.  Sometimes, due to
measurement error, it is the third point in the flare event which is
the highest and then the flare fit chi-squared is bad.  With only 4 points
total, there are not enough points to compensate for this one bad point
and the microlensing fit will be better than the flare fit.
Thus, due to random measurement errors, 
bumps that are near our minimum of a total of 4 flux points are sometimes
fit better with a microlensing shape, even when they are probably 
flare events.  

To solve this problem, we up our required minimum number of points from 4 to 5
sequential 3-sigma high measurements and calculate an additional
asymmetry statistic, which distinguishes the symmetric microlensing shape
from the asymmetric flare shape.

To calculate {\bf asymmetry} we start by finding a symmetry timescale
which is the larger of 1.5 times the flare fit timescale $\lambda$ and
twice the microlensing timescale $\hat t$.
We define {\bf asymmetry} as the sum of absolute values of the 
differences between the flux measured at a time before the peak
and the flux measured at the symmetric time after the peak, all
normalized to the total flux above the median under the peak 
(See Table~\ref{tab:cuts} for the formulas).
Requiring that {\bf asymmetry} be smaller than 17\%, together
with the above chi-squared criterion, and the new minimum number of data points
effectively remove the flare events, including those of short duration.

We also implement two more signal-noise criteria, first by requiring that 
{\bf mlchi2dof} (the chi-squared per degree of freedom of the microlensing fit defined above) 
be less than 3.5.  This ensures that the microlensing shape is a relatively good fit to theory.
The bottom lightcurve in Figure~\ref{fig:badlc} shows an example bump removed
by the chi-squared criteria.
We also compare {\bf chi2in}, the chi-squared per dof of the fit under the 
peak, with {\bf chi2out}, the chi-squared
per dof outside the peak.  We define the a region centered on the peak of duration
3 times the event duration ${\hat t}$ as ``under the peak" and a region 6 times 
the peak duration that excludes the above region as ``outside the peak".  
Our selection criterion is to
require that the ratio of {\bf chi2in} to {\bf chi2out} be less than 4.
This selection criterion eliminates rare bumps where the extra noise in the peak area causes
our bump criteria to be met.  We don't want bumps where the fit chi-squared
under the peak is much worse than the chi-squared outside the peak.

\subsubsection{Foreground Moving Objects in the Kepler field}
\begin{figure}[htb!]
\begin{center}
\includegraphics[scale=.3,angle=-90,trim=0.5in 0 0 0]{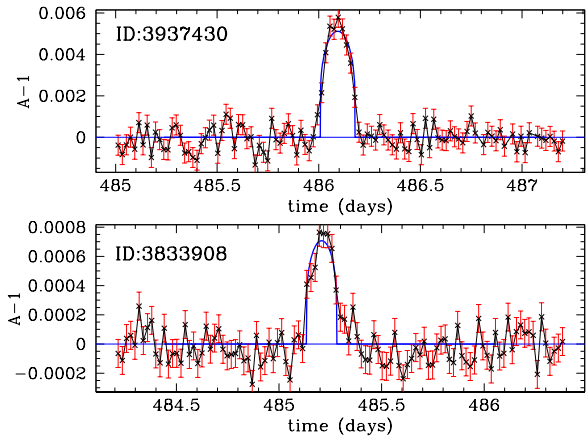}
\caption{
Examples of bumps in Kepler lightcurves caused by comet C/2006 Q1 (McNaught) during quarter 5. 
The Kepler source star IDs are shown and the solid blue line is 
a fit microlensing model.} 
\label{fig:cometlc5}
\end{center}
\end{figure}

\begin{figure}[htb!]
\begin{center}
\includegraphics[scale=.3,angle=-90,trim=0.5in 0 0 0]{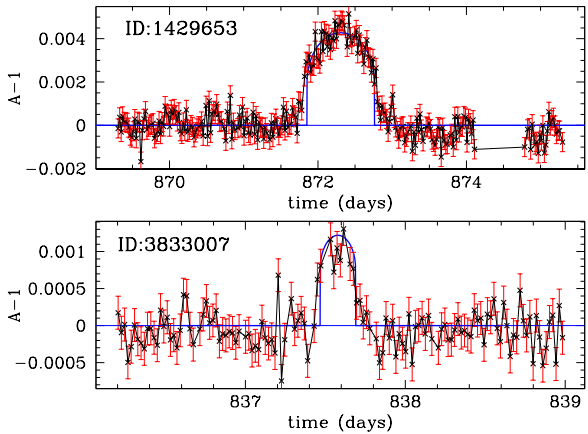}
\caption{
Examples of bumps in Kepler lightcurves caused by comets during quarter 9.  
The upper panel is from C/2007 Q3  (see Table~\ref{tab:comets}) and 
the lower panel from an unidentified comet or asteroid.
The Kepler source star IDs are shown as are fits to a microlensing shape.} \label{fig:cometlc9}
\end{center}
\end{figure}

After applying these cuts to 2 years of data we find 17 candidate microlensing
events.  These are listed in Table~\ref{tab:comets} and examples are shown in 
Figures~\ref{fig:cometlc5} and \ref{fig:cometlc9}.

\begin{figure}[htb!]
\begin{center}
\includegraphics[scale=.3,angle=-90,trim=0.5in 0 0 0]{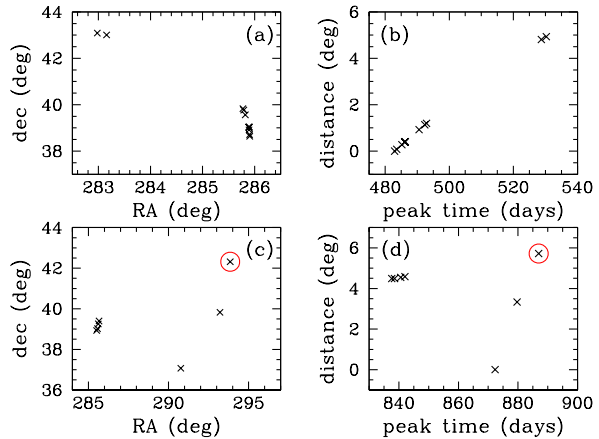}
\caption{
Part (a) shows right ascension (RA) and declination (dec) of bumps
caused by comet C/2006 Q1 as it passed through the
Kepler field during quarter 5.  
Part (b) shows the angular distance moved by the comet vs. Kepler time. 
Part (c) shows RA and dec of two separate objects moving through the field 
during quarter 9, and part (d) shows the distance moved by these two objects.  
The constant slopes in parts (b) and (d) imply a constant angular
speed.  The point circled is a bump that did not pass all the 
cuts but is shown here so there are more than two points on the line.
\label{fig:comets}}
\end{center}
\end{figure}

Note that these events are nicely symmetric and reasonable fits to the
microlensing (actually just stellar limb darkening) shape.  
However, the candidate events only occur in quarters 5 and 9, and as shown
in Figure~\ref{fig:comets} occur in a pattern on the sky
that makes it extremely unlikely that they
are due to microlensing.  Considering first the quarter 5 events as shown in 
Figure~\ref{fig:comets} (a) and (b),
we see a clear track across the Kepler field. 
From the times shown in Figure~\ref{fig:comets} (b) we see
that an object entered the
Kepler field near the lower right hand corner (Figure~\ref{fig:comets} (a))
and moved at a constant rate to the upper left hand corner over a period of around
60 days.  From the times of the bump peaks as given 
in Table~\ref{tab:comets}, it is clear that this is an object that moved through the
Kepler field leaving a track of transient brightenings in its path.
Note that the probability of microlensing is so small that it 
is extremely unlikely that these
events are all caused by microlensing by a single lens.

From the time differences between the bump peaks we deduce an angular
speed of around 16 arcsec/hr.  Assuming the motion is due to the
30 km/s Kepler satellite motion around the Sun, 
this implies a distance to this object of around 10 AU.
We thus expect all these quarter 5 events to be caused by a comet passing
through the Kepler field.  

Using the fractional magnification from the peak flux
of each event, and the g magnitude of each Kepler source star (given in the Kepler
lightcurve file header), we can
calculate a magnitude that needs to be added to each star to explain each 
bump.  We find values ranging from gmag = 20 to gmag=21.5.  The largest of
these values gives a minimum magnitude of the comet.

Using the right ascension (RA) and declination (dec) of the Kepler 
stars and making the coordinate
transformation from the Earth trailing Kepler orbit frame to the 
Earth frame we examine objects the using the Minor Planet Center (2013)
software and find that comet C/2006 Q1 (McNaught)
passed through the locations of our quarter 5 candidate events at the times they occurred.
This comet was at a distance of around 10 AU, just as we estimated.
Thus all these quarter 5 candidate microlensing events are 
consistent with being caused by comet C/2006 Q1.

Similarly examining the quarter 9 candidate events as shown 
in Figures~\ref{fig:cometlc9} and \ref{fig:comets},
we see evidence for two tracks that cross the field.  Using the time
between bumps (Table~\ref{tab:comets}) we again conclude that these events are caused by
two objects moving through the Kepler field at a constant angular speed.
For one of the new objects there were only 2 bumps that passed all our cuts,
which does not make for much of a test for constant 
motion, so we loosened the cuts somewhat
and found more events along this track.  We show one such additional bump as a 
circled point in Figure~\ref{fig:comets}(c) and (d).
Using the positions and times we were able to identify one of these objects
as comet C/2007 Q3.
The other object is moving faster than the two comets and so is
probably closer.  We could not identify a counterpart
in the Minor Planet Ephemeris (Minor Planet Center, 2013), so it may be a previously unidentified comet or
asteroid.
Using the same method as for C/2006 Q1 we find a minimum magnitudes of around
$\mathrm{gmag}=21.3$ for C/2007 Q3, and $\mathrm{gmag}=21$ for the new bright object.
In any case, we remove these events as microlensing candidates.
\begin{deluxetable}{ccccccc}
\tabletypesize{\scriptsize}
\tablecaption{Comets in Kepler data \label{tab:comets}}
\tablewidth{0pt}
\tablehead{
\colhead{quarter} &
\colhead{Kepler ID} &
\colhead{time$^{\rm a}$}  &
\colhead{RA$^{\rm b}$} &
\colhead{dec$^{\rm b}$} &
\colhead{gmag$^{\rm c}$} &
\colhead{comet}
}
\startdata
5& 3527753  & 482.949  & 285.905 &38.6395  & 21.3 & C/2006 Q1\\
5& 3628766 & 483.582  & 285.904 & 38.7161 & 21.4 & C/2006 Q1\\
5& 3833908 & 485.217  & 285.893 & 38.9171 & 20.8 & C/2006 Q1\\
5& 3937408 & 486.116  & 285.887 & 39.029  & 21.5 & C/2006 Q1\\
5& 3937430 & 486.106  & 285.894 & 39.0263 & 21.2 & C/2006 Q1\\
5& 3937432 & 486.259  & 285.894 & 39.0463 & 21.4 & C/2006 Q1\\
5& 4447346 & 490.591  & 285.82  & 39.5627 & 21.7 & C/2006 Q1\\ 
5& 4637389 & 492.379 & 285.791 & 39.7713 & 22.1 & C/2006 Q1\\
5& 4729654 & 492.880 & 285.774 & 39.8299 & 21.9 & C/2006 Q1\\
5& 7421340 & 530.305 & 282.978 & 43.0877 & 20.8 & C/2006 Q1\\
5& 7421791 & 528.762 & 283.157 & 43.0068 & 20.6 & C/2006 Q1\\
9& 1429653 & 872.332 & 290.769 & 37.0712 & 21.3 & new \\
9& 3833007 & 837.604 & 285.517 & 38.9324 & 21.9 & C/2007 Q3 \\
9& 3936698 & 838.523 & 285.568 & 39.023 & 21.5 & C/2007 Q3 \\
9& 4138614 & 840.608 & 285.615 & 39.2529 & 21.5 & C/2007 Q3\\
9& 4347043 & 842.049 & 285.662 & 39.4001 & 22.6 & C/2007 Q3\\
9& 4751561 & 879.688 & 293.212 & 39.8243 & 21.2 & new \\
9& 6870049$^*$ & 886.968 & 293.851 & 42.3156 &  21.9 & new \\
\enddata
\tablenotetext{a}{Kepler days; convert to Julian date by adding 2454833}
\tablenotetext{b}{From the Kepler satellite point of view}
\tablenotetext{c}{Found by adding the bump peak flux to the source star magnitude. See text.}
\tablenotetext{*}{Did not pass all cuts.}
\end{deluxetable}

After removing the events due to bright moving objects
from our candidate list we are left with no PBH microlensing candidates.

\section{Efficiency Calculation}
Since we found no candidates, we can place upper limits on the halo density
of PBH DM.  We display these as limits on the halo fraction, where 
a dark matter halo made entirely of PBH DM and a local density of 
$\rho_\mathrm{DM}=0.3$ GeV${\rm cm}^{-3}$ ($0.0079 M_\odot{\rm pc}^{-3}$)
would give a halo fraction of unity.
In order to do this we need to calculate the number of microlensing events
our search through the Kepler data would be expected to detect if the halo
consisted entirely of PBH DM.
We did this estimate previously (Paper I and Paper II), but in those
theoretical analyses we made several approximations that our actual search
through the data has not substantiated.  For example, we assumed Gaussian
errors while the actual data has many non-Gaussian excursions in flux
resulting in many noise events.  We did not know about the flare events 
or the frequency of other variability induced events.  Thus we need to
recalculate the expected number of detections using the actual selection
criteria that gave us no PBH DM candidates.
We use the same halo model as in Paper I and Paper II:  a constant halo density
between the Earth the Kepler field stars about 1 kpc away in a direction almost
orthogonal to the Galaxy center, an isotropic Maxwellian velocity distribution,
and no need for any solar motion transverse to the line-of-sight since the Sun's
rotation around the Galaxy center is in the direction of the Kepler field stars.

Using this model, we calculate the expected number of event detections
by constructing simulated PBH microlensing events and adding these
into actual Kepler lightcurves.  We then analyze these simulated events
using the same software and selection criteria that we used for 
our microlensing search (including, of course, removal of the ``bad data" ranges
of the lightcurves).  To create the simulated microlensing lightcurves
we use the full limb-darkened microlensing formula (Witt \& Mao 1994, corrected in
equation~13 of Cieplak \& Griest (2013)) with linear limb darkening coefficients
calculated using the Sing (2010) model grid for each Kepler source star (Paper II).

In the Monte Carlo efficiency calculation we want to cover all possible actual microlensing events, so we add
simulated events covering the possible range of physical parameters.  These are
discussed in great detail in Papers I and II, but include $m$, the DM mass,
$t_0$ the time of closest approach between the lens and source lines-of-sights,
$x$, the distance to the lens divided by the distance to the source star,
$v_t$, the transverse speed of the lens relative to the source line-of-sight,
and $u_{min}=ux/r_E$, distance of closest approach scaled by the Einstein radius in the lens plane. 
The Einstein radius is given by
\begin{equation}
r_E=0.0193 \sqrt{x(1-x)}[(L/{\rm kpc})\mn]^{1/2}\rsun, 
\end{equation}
where $\mn=(m/\aten{-9}\msun)$ and $L$ is the source star distance.

Each time a simulated event is detected we calculate the differential
microlensing rate for the microlensing parameters used, 
and for that specific source star,
using Equation 19 of Cieplak \& Griest (2013):
\begin{equation}
 \frac{d\Gamma}{dx du_{\mathrm{min}} dv_\mathrm{t}} = 4 r_\mathrm{E}(x) L \frac{\rho}{M} \frac{v_\mathrm{t}^2}{v_\mathrm{c}^2} e^{-v_\mathrm{t}^2/v_\mathrm{c}^2}.
\end{equation}
where
$v_c = 220$km/s is the assumed circular speed around the Milky Way at the Sun's position.
Note, $v_c$ sets the dispersion in the assumed isotropic Maxwellian velocity distribution of the
dark matter halo. 

One weakness in the above calculation is our estimate of the source star distance.
The header of each Kepler lightcurve file contains
the stellar radius, $R_*$,
Sloan $r$ and $g$ magnitudes, effective temperature, $T_\mathrm{eff}$,
star position (RA and dec),
extinction parameters $A_V$ and $E(B-V)$, etc.
We estimate the apparent visual magnitude from:
\begin{equation}
V = g - 0.0026 - 0.533(g-r)
\end{equation}
(Fukugita, et al. 1996), and the stellar distance from:
\begin{equation}
L \approx 1.19\ten{-3} R_* (T_\mathrm{eff}/T_\odot)^2
\aten{0.2(V-A_V+{\rm B.C.})}\mathrm{kpc},
\end{equation}
where B.C. is the bolometric correction.
Note that we make a crude bolometric correction, using
only the effective temperature and whether the source is a main sequence
or giant star (Carroll \& Ostlie 2007),
but we include it because it slightly reduces the distances to the sources,
thereby reducing the expected detection rate, and we want
our calculation to be conservative.

By adding many millions of simulated events covering the entire
allowed range of microlensing parameters we effectively perform
the efficiency weighted integral over the above differential rate 
(Equation~2), where $\Gamma$ is in units of expected events per day per star.
By multiplying this by the number of non-variable stars and the duration of
the lightcurves we thus find the
efficiency-corrected number of expected microlensing events.  
The number of non-variable stars and star-days for each Kepler quarter
is given in Table~\ref{tab:stardays}.

\begin{deluxetable}{ccccccc}
\tabletypesize{\scriptsize}
\tablecaption{Star-days by Kepler Quarter\label{tab:stardays}}
\tablewidth{0pt}
\tablehead{
\colhead{quarter} &
\colhead{total stars} &
\colhead{non-variable stars} &
\colhead{start time$^{\rm a}$} &
\colhead{end time$^{\rm a}$} &
\colhead{duration (days)} &
\colhead{star-days}
}
\startdata
2 &  152238 & 94758 &  169.765 & 258.467 & 88.70 &   23011 \\
3 &  152453 & 99528 &  260.244 & 349.495 & 89.25 &   24320 \\
4 &  156994 & 80589 &  352.37  & 442.20  & 89.83 &   19842 \\
5 &  152376 & 102323 &  443.50 & 538.16 &  94.66 &   26519 \\
6 &  145973 & 96268 &  539.47  & 629.30  & 89.83 &   23676 \\
7 &  151345 & 66468 &  630.20  & 719.55  & 89.25 &   16242 \\
8 &  154640 & 84670 &  735.38  & 802.34  & 66.96 &   15522 \\
9 &  150647 & 73856 &  808.52  & 905.93  & 97.41 &   19697 \\
\enddata
\tablenotetext{a}{Kepler days; convert to Julian date by adding 2454833}
\end{deluxetable}

Since we detected no events, a
95\% C.L. upper limit of $3/N_\mathrm{exp}$ can be set on the halo fraction.
This is a similar to the method used previously by the MACHO collaboration
(Alcock, et al. 1996) to set limits on low mass MACHO DM.  
Since the number density of PBHs, the efficiency, and the differential rate all
depend strongly on the assumed PBH mass, we perform this Monte Carlo calculation independently for each
value of PBH mass.  Also, since each quarter of Kepler lightcurve data
includes somewhat different source stars and has different noise characteristics
we also calculate the expected number of events independently for each
quarter and sum the total to find our limits. 
All together we analyzed more than 500 million simulated events. 
Our results are plotted in Figure~\ref{fig:halofrac}.

\section{Limits on PBD Dark Matter and Discussion}
\begin{figure}[htb!]
\begin{center}
\includegraphics[scale=.3,angle=-90,trim=0.5in 0 0 0]{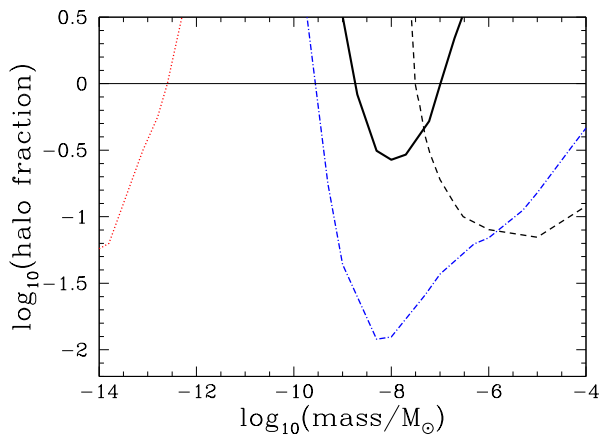}
\caption{Upper limits (95\% C.L.) on PBH DM from non-observation of PBH microlensing in two years
of Kepler data.  The solid black line is our new limit, the dashed black line is the previous best limit
(Alcock, et al. (1998)), the blue dot-dash line is the theoretical limit from Paper II, and the
red dotted line is the femtolensing limit from Barnacka, et al. (2012). 
The black horizontal line indicates a halo
density of 0.3 GeV${\rm cm}^{-3}$.
} \label{fig:halofrac}
\end{center}
\end{figure}

The thick black solid line in 
Figure~\ref{fig:halofrac} shows our new limits on the possibility of PBH DM.  
Since these limits depend only on the mass (assuming the lens is a compact
object) they apply to any massive compact halo object, so are robust
limits on planets, non-topological solitons, etc.
Our analysis shows that PBH DM with 
masses in the range $2\ten{-9}\msun$ to $\aten{-7}\msun$ 
cannot make up the entirety of the dark matter in a canonical DM halo.

Also shown in Figure~\ref{fig:halofrac} is a black dashed line which shows the best previous
limits from a combined MACHO-EROS analysis in 1998 (Alcock, et al. 1998).  
Note some authors (e.g. Carr, et al. 2010) only quote the 2007 EROS-only limits 
(Tisserand, et al. 2007) but these are not as strong as the earlier combined
limits.  We see that our new limits rule out more than one order of magnitude
more of the allowed PBH DM mass range, for the first time eliminating masses between
$2\ten{-9}\msun$ and $3\ten{-8}\msun$ as being the entirety of the DM.

Also shown in Figure~\ref{fig:halofrac} as a blue dot-dash line are the potential theoretical 
limits from analysis of the entire 8-years of Kepler data from Paper II.
Naively we would have expected our limits to be about 1/4 of these theoretical
limits since we analyzed 2 years of data, about 1/4 of the total 
of 8 years of data assumed in Paper II.
The figure shows our experimental limits are about a factor of 8 weaker
than expected.  This is due to the overly optimistic efficiency assumptions made in
Papers I and II.  Because of the existence of many flare events, including those
of short duration,
we had to increase the required number of sequential high points from 4 to 5.
Also non-Gaussian noise and systematic errors in the data meant
we needed to add several other shape and signal-to-noise selection 
criteria, all of which reduced the number of microlensing events we expect
to find.  

Finally, the dotted red line at the left of Figure~\ref{fig:halofrac} shows the
recent femtolensing limits from Barnacka, et al. (2012), which define the lower edge of the PBH
allowed mass range.  We see that there are still about 4 orders of magnitude
in mass (from $3\ten{-13}\msun$ to $2\ten{-9}\msun$) where PBH DM (or MACHO DM) can make up the entirely
of the DM.  Future analysis of the entire Kepler data set should discover PBH DM
or eliminate some portion of this range, and future missions such as WFIRST
have the potential to cover another order of magnitude (Paper II).

\acknowledgments

K.G. and A.M.C. were supported in part by the U.S. Department of Energy
under grants DE-FG03-97ER40546 and DE-SC0009919. 
A.M.C. was supported in part by the National Science Foundation Graduate Research Fellowship under grant number DGE0707423. 
Some of the data presented in this paper were obtained from the Multimission Archive at the Space Telescope Science Institute (MAST). STScI 
is operated by the Association of Universities for Research in Astronomy, Inc., under NASA contract NAS5-26555. Support for MAST for non-HST data is provided by the NASA Office of Space Science via grant NNX09AF08G and by other grants and contracts.




\bibliographystyle{apj}
\smallskip
\centerline{\bf REFERENCES}
\smallskip

\noindent
Aad, G. et al., 2012, Phys Lett. B, 716,1

\noindent
Ade, P.A.R., et al., 2013, arXiv:1303:5076 

\noindent
Alcock, C., et al., 1996, Astrophys. J. \textbf{471}, 774 

\noindent
Alcock, C., et al., 1998, Astrophy J. Lett. {\bf 499}, L9 

\noindent
Alcock, C., et al., 2000, Astrophys. J. \textbf{542}, 281 

\noindent
Alcock, C., et al., 2001, Astrophy J. Lett. {\bf 550}, L169 

\noindent
ATLAS Collaboration, 2013, ATLAS-CONF-2013-001

\noindent
Barnacka, A., Glicenstein, J.F., \& Moderski, R., 2012, Phys. Rev. D, 86, 043001

\noindent
Borucki, et al., 2010, Science, {\bf 327}, 977

\noindent
Carr, B.J., Kazunori, K., Sendouda, Y., \& Yokoyama, J., 2010, Phys Rev. D81, 104019 

\noindent
Carroll, B.W. \& Ostlie, D.A., 2007,
  ``An Introduction to Modern Astrophysics", 2nd Edition, Appendix G,
   Person/Addison Wesley

\noindent
Chatrchyan, S. et al., 2013, Phys. Lett. B, 716, 30

\noindent
Chatrchyan, S. et al., 2012, Phys. Rev. Lett. 109,171803 

\noindent
Ciardi, D. R., el al., 2011, AJ, 141, 108

\noindent
Cieplak, A.M. \& Griest, K., 2013, ApJ, 766, 145 {\bf (Paper II)}

\noindent
CMS Collaboration, 2013, CMS PAS SUS-12-023; CMS PAS SUS-12-028

\noindent
Feng, J.L., 2010, ARA\&A, 48,495

\noindent
Frampton, P.H., et al., 2010, JCAP {\bf 04}, 023

\noindent
Fraquelli, D. \& Thompson, S.E., 2011, Kepler Archive Manual, KDMC-10008-002, (http://archive.stsci.edu/kepler)

\noindent
Fukugita, M., et al., 1996, AJ, bf 111, 1748 

\noindent
Griest, K., 1991, Astrophys. J. {\bf 366}, 412 

\noindent
Griest, K., Cieplak, A.M., \& Lehner, M.J., 2013, submitted.

\noindent
Griest, K., Lehner, M.J., Cieplak, A.M., \& Jain, B., 2011, Phy. Rev. Lett. 107, 231101 {\bf (Paper I)}

\noindent
Jungman, G., Kamionkowski, M, \& Griest, K., 1996, Phys. Rep., 267,195

\noindent
Kawasaki, W., Sugiyama,  N., \&Yanagida, T., 1998, Phys. Rev. D 57, 6050 

\noindent
Koch, D. G., et al. 2010, ApJL,, 713, L79

\noindent
Khlopov, M.Y., 2008, in Recent Advances on the Physics of compact objects and Gravitational Waves, edited by J. A. de Freitas Pacheco

\noindent
Martin, S.P., 2011, arXiv:hep-ph/9709356v6 

\noindent
Minor Planet Center, 2013, http://www.minorplanetcenter.net

\noindent
Paczynski, B, 1986, Astrophys. J. {\bf 304}, 1

\noindent
Sing. D.K., 2010, A\&A. 510. A21

\noindent
Tisserand, P., et al.,  2007, Astron. \& Astrophys. {\bf 469}, 387

\noindent
Witt, H.J. \& Mao, S. 1994, ApJ, 430, 505


\end{document}